\documentclass{article}

\usepackage{cite} %
\usepackage{amsmath,amssymb} %
\usepackage{graphicx,epstopdf} %
\usepackage{times,fullpage} %
\usepackage{longtable}

\IfFileExists{srcltx.sty}{\usepackage[active]{srcltx}}

\usepackage[a4paper,centering,hscale=0.7,vscale=0.7]{geometry}

\usepackage{bm,paralist,xspace,array}

\usepackage[breaklinks=true]{hyperref} %

\IfFileExists{hypernat.sty}{\usepackage{hypernat}}

\begin{document}

\title{New evidence for dark matter} %

\author{A.~Boyarsky$^{1,2}$, O.~Ruchayskiy$^1$,  D.~Iakubovskyi$^2$, A.V.~Macci\`o$^3$, D.~Malyshev$^{4}$\\
  \normalsize{$^{1}$Ecole Polytechnique F\'ed\'erale de Lausanne,}\\
  \normalsize{FSB/ITP/LPPC, BSP CH-1015, Lausanne, Switzerland}\\
  \normalsize{$^{2}$Bogolyubov Institute for Theoretical Physics,}\\
  \normalsize{Metrologichna str., 14-b, Kiev 03680, Ukraine}\\
  \normalsize{$^3$Max-Planck-Institut f\"ur Astronomie, K\"onigstuhl 17, 69117
    Heidelberg, Germany}\\
  \normalsize{$^4$Dublin Institute for Advanced Studies, 31 Fitzwilliam
    Place, Dublin 2, Ireland} \\
} \date{} %

\newcommand{\dm}{{\textsc{DM}}} 
\newcommand{\cdm}{{\textsc{cdm}}} %
\newcommand{\lcdm}{\ensuremath{\Lambda}\textsc{CDM}\xspace} %
\newcommand{\cwdm}{{\textsc{cwdm}}} %
\newcommand{\wdm}{\textsc{wdm}} %
\newcommand{\fwdm}{{\ensuremath{\bm{f}_\wdm}}\xspace}%
\newcommand{\fov}{{\mathrm{fov}}} 
\newcommand{\eV} {\ensuremath{\:\mathrm{eV}}}   %
\newcommand{\keV}{\ensuremath{\:\mathrm{keV}}} %
\newcommand{\cm}{\:\mathrm{cm}} 
\newcommand{\s}{\:\mathrm{sec}} 
\newcommand{\ks}{\:\mathrm{ksec}} 
\newcommand{\sr}{\:\mathrm{sr}} 
\newcommand{\ph}{\:\mathrm{ph}} 
\newcommand{\pc}{\:\mathrm{pc}} 
\newcommand{\kpc}{\:\mathrm{kpc}} 
\newcommand{\Mpc}{\:\mathrm{Mpc}} 
\newcommand{\mpc}{\:\mathrm{Mpc}} 
\newcommand{\Gpc}{\:\mathrm{Gpc}} 
\newcommand{\g}{\:\mathrm{g}} 
\newcommand{\cts}{\:\mathrm{cts}} 
\newcommand{\xmm}{\textsl{XMM-Newton}\xspace} %
\newcommand{\nfw}{\textsc{nfw}} %
\newcommand{\iso}{\textsc{iso}} %
\newcommand{\burk}{\textsc{burk}} %
\renewcommand{\S}{\mathcal{S}} %
\newcommand{\totalprof}{1095\xspace} %
\newcommand{\totalprofsel}{805\xspace}%
\newcommand{\nfwprofsel}{490\xspace}%
\newcommand{\isoprofsel}{285\xspace}%
\newcommand{\burkprofsel}{30\xspace}%
\newcommand{\total}{357\xspace}
\newcommand{\totalsel}{289\xspace}
\newcommand{\clusters}{130\xspace}%
\newcommand{\clusterssel}{121\xspace}
\newcommand{\groups}{26\xspace} %
\newcommand{\groupssel}{25\xspace} %
\newcommand{\spirals}{180\xspace}%
\newcommand{\spiralssel}{124\xspace}%
\newcommand{\ellipticals}{10\xspace} %
\newcommand{\ellipticalssel}{10\xspace} %
\newcommand{\dsphs}{11\xspace} %
\newcommand{\dsphssel}{11\xspace} %
\newcommand{\nfwiso}{76\xspace} 
\newcommand{\nfwisosel}{59\xspace} 
\newcommand{\rmaxsel}{891} 


\baselineskip12pt

\maketitle

\textbf{Observations of star motion, emissions from hot ionized gas,
  gravitational lensing and other tracers demonstrate that the dynamics of
  galaxies and galaxy clusters cannot be explained by the Newtonian potential
  produced by visible matter
  only\cite{Evrard:95,Buote:03,Refregier:03,Bergstrom:00}. The simplest
  resolution assumes that a significant fraction of matter in the Universe,
  dominating the dynamics of objects from dwarf galaxies to galaxy clusters,
  does not interact with electromagnetic radiation (hence the name \emph{dark
    matter}).  This elegant hypothesis poses, however, a major challenge to
  the highly successful Standard Model of particle physics, as it was realized
  that dark matter cannot be made of known elementary
  particles\cite{Bergstrom:00}. The quest for direct evidence of the presence
  of dark matter and for its properties thus becomes of crucial importance for
  building a fundamental theory of nature.  Here we present a new universal
  relation, satisfied by matter distributions at all observed scales, and show
  its amazingly good and detailed agreement with the predictions of the most
  up-to-date \emph{pure dark matter} simulations of structure formation in the
  Universe\cite{Neto:07,Maccio:08,Springel:08a}.  This behaviour seems to be
  insensitive to the complicated feedback of ordinary matter on dark matter.
  Therefore, it potentially allows to compare theoretical predictions directly
  with observations, thus providing a new tool to constrain the properties of
  dark matter.  This work extends the previous
  analysis\cite{Kormendy:04,Donato:09,Gentile:09a} to a larger range of
  masses, demonstrates a different scaling law, and compares it with numerical
  simulations.  Such a universal property, observed in structures of all sizes
  (from dwarf spheroidal galaxies to galaxy clusters), is difficult to explain
  without dark matter, thus providing new evidence for its existence.  }

Current cosmological data, including the earliest stage of structure
formation, are well-described by the ``concordance model''
($\Lambda$CDM)\cite{WMAP5cosmoParams}.  The microscopic properties of the dark
matter particles can strongly affect the formation of cosmic structures,
originating from tiny preturbations in the primordial density field. Within
the $\Lambda$CDM the dark matter is assumed to be non-interacting and to have
no primordial velocity -- \emph{cold dark matter} (CDM).  To describe the
formation and properties of virialized DM objects at a later, non-linear,
stage of structure formation, numerical N-body simulations have been
extensively used\cite{Bertschinger:98}. It was demonstrated that within the
CDM model spherically averaged density profiles of DM halos of all scales are
self-similar and universal\cite{Navarro:96,Stadel:08}.

Many works have addressed the issue of measuring the DM distribution in
observed objects (see
e.g.\cite{Evrard:95,Buote:03,Refregier:03,Burkert:95,Gilmore:07a,Gentile:07b}).
Two classes of profiles have been used to fit observational data: profiles
with singular, cusp-like behaviour at small radii $r$~(e.g. NFW
profile\cite{Navarro:96}) or profiles that tend to a constant central density
$\rho_c$ -- \emph{cored profiles}, such as e.g.  pseudo-isothermal
profile\cite{Kent:86} (ISO) or Burkert profile\cite{Burkert:95} (BURK).  Both
profile families are described by two basic parameters: a characteristic scale
$r_\star$ at which the inner slope of the DM density profile changes towards
its outer asymptotic and an average density~$\rho_\star$ within this radius.

Unfortunately current data do not allow us to determine, in
a conclusive way, the presence of cores or cusps in the observed density
distributions. Often the same data set can be equally well fitted by profiles
of different
type\cite{Goerdt:06,Sanchez:06b,Gilmore:07a,Gentile:07b,deBlok:09}.  The
deviation from predictions of pure CDM models can occur for several reasons:
for example, a significant influence of baryons; the microscopic properties of
dark matter particles different from those of CDM (non-gravitational
self-interaction or interaction with baryons, non-zero primordial velocities).
Moreover baryons can contribute significantly (and may even dominate) the
total mass profile in the inner regions, and thus pure CDM predictions are
difficult to test.

Recently the works\cite{Donato:09,Gentile:09a} analysed rotation curves and
weak lensing data for a sample of dwarf, spiral and elliptical galaxies fitted
by the Burkert profile and demonstrated that the mean dark matter surface
density $\langle\Sigma\rangle =\rho_\star r_\star$ remains constant for all
these galaxies.  In this work we extend the analysis
of\cite{Donato:09,Gentile:09a} to galaxies and galaxy clusters.  We have
complied a catalog of DM distribution in various celestial objects from more
than 50 scientific publications. After applying uniform selection criteria to
this catalog (see \textbf{Method summary}), we were left with \totalprofsel
Dark Matter profiles (\nfwprofsel NFW, \isoprofsel ISO and \burkprofsel BURK)
from \totalsel unique objects: \spiralssel spiral galaxies, \dsphssel dwarf
spheroidals (dSphs), \ellipticalssel elliptical galaxies, \groupssel galaxy
groups and \clusterssel galaxy clusters.  To properly compare DM
distributions, fitted by different density profiles, we introduce a \emph{dark
  matter column density}, averaged over the central part of an object:
\begin{equation}
 \label{eq:Sbar}
 \S = \frac2{r_\star^2}\int^{r_\star}_0 rdr\, \hskip -1ex
 \int  dz\, \rho_\dm(\sqrt{r^2+z^2})
 \simeq \frac{M_\text{cyl}}{\pi r_\star^2}\;,
\end{equation}
Integral over $z$ extends to the virial boundary of a DM halo. The
definition~(\ref{eq:Sbar}) implies that $\S$ is proportional to the dark
matter surface density within $r_\star$ ($\S\propto \rho_\star r_\star$).  For
distant objects $\S$ is defined via $M_\text{cyl}$ -- mass within a cylinder
of radius $r_\star$.  Parameters of different profiles that fit the same DM
density distribution are related (for example, $r_s$ for NFW is equal to
$6.1r_c$ for ISO and equals to $1.6r_B$ for BURK). Choosing these values as
$r_\star$ in each case, one finds that the value of $\S$ for NFW (cuspy
profile with $r^{-3}$ asymptotic at large radii) and ISO (cored profile with
$r^{-2}$ behaviour at large scales) differ by less than 10\% (the difference
between NFW and BURK is $\sim 2\%$), see Supporting Information for details.
Thus, the DM column density $\S$ is \emph{insensitive to the type of DM
  density profile, used to fit the same observational data}.

Our final dataset spans more than four orders of magnitude in $r_\star$
($0.2\kpc\lesssim r_\star \lesssim 2.5\mpc$) and about eight orders of
magnitude in halo masses.\footnote{We use $M_{200}$ as halo mass
  $M_\text{halo}$. A proper definition of $M_{200}$ can be found e.g.
  in\cite{Maccio:08} or in the Supplementary Information.}  The observational
data can be fit by a single power-law:
\begin{equation}
\lg \S = 0.21 \lg
\frac{M_\text{halo}}{10^{10}M_\odot} + 1.79\label{eq:2}
\end{equation}
(with $\S$ in $M_\odot \pc^{-2}$).  One could try to interpret the data
presented in Fig.~1 in spirit of\cite{Kormendy:04,Donato:09,Gentile:09a} (i.e.
$\S=\mathrm{const}$), although with a higher value of $\S \approx 260
M_\odot\pc^{-2}$ (as was originally suggested in\cite{Boyarsky:06c}). The
apparent trend at higher masses could then be attributed to systematic errors
and observational bias.  Indeed, the results are based on observational data
of different nature and different quality and we do not consider in this work
observational errors.  However, both $M_\text{halo}$ and $\S$ were computed,
using 3 density profiles on average, which should diminish the errors. The
scaling relation~(\ref{eq:2}) is supported by the data, spread over many
orders in magnitude in halo mass, and the shift of some data points should not
significantly affect the systematic trend.

Next, we compare our data with outcomes from cosmological N-body simulations
within the $\Lambda$CDM model\cite{Maccio:08,Springel:08a}.  For each
simulated halo we compute $M_\text{halo}$, fit the particle distribution to
the NFW density profile and calculate $\S$ using formula~(\ref{eq:Sbar}). The
observational data together with results from $\Lambda$CDM numerical
simulations\cite{Maccio:08} is plotted of the Fig.~1.  The black dashed-dotted
line on this Figure is the $\S-M_\text{halo}$ relation obtained from N-body
simulations\cite{Maccio:08}, using the WMAP fifth year\cite{WMAP5cosmoParams}
cosmological parameters. It fairly well reproduces the fit~(\ref{eq:2}).
Moreover, the pink shaded region (showing the $3\sigma$ scatter in the
simulation data) contains most of the observational points within the halo
mass range, probed by simulations.  Therefore, the observed $M_\text{halo}-\S$
scaling coincides with the relation between the parameters of DM density
profiles observed in numerical simulations for long
time\cite{Navarro:96,Neto:07,Maccio:08} over more than five orders of
magnitude in mass.

Dwarf spheroidal satellites (dSphs) of the Milky Way (orange diamonds on the
Fig.~1) do not follow the relation~(\ref{eq:2}). Recently the Aquarius project
has produced a statistically significant sample of well resolved density
profiles for satellite halos\cite{Springel:08a}, making it possible to
determine their $r_\star-\rho_\star$ relation.  Satellites were found to be
more concentrated than isolated halos and thus have a higher value of $\S$ at
fixed $M_\text{halo}$.  Fig.~1 shows that the $\S-M_\text{halo}$ relation for
satellite halos (gray dashed line) from the Aquarius
simulation\cite{Springel:08a} reproduces well the data on dSphs.

The fit to the data without the dSphs has the slope $\approx 0.23$, much
better quality of fit, and coincides extremely well with the results of N-body
simulations\cite{Maccio:08} for isolated halos (black dashed-dotted line on
Fig.~1). At masses below $10^{10}M_\odot$ no isolated halos were resolved
in\cite{Maccio:08} and a simple toy model\cite{Bullock:99,Maccio:08} was used
to predict the relation between parameters of NFW profile in a given
cosmological scenario.  The model (dotted line in the Fig.~1) fits well the
results for the few spiral galaxies in this range. \emph{Thus the agreement
  between observations and predictions from $\Lambda$CDM extend over more than
  eight orders of magnitude in mass}.

Comparison of our data with theoretical predictions (N-body simulations in our
case) indicates that, despite the presence of various systematic errors in the
data, the DM distributions in the observed objects exhibit a universal
property -- a systematic change of the average column density $\S$ as a
function of the object mass ($\S \propto M_\text{halo}^{0.2}$,
relation~(\ref{eq:2})). This is different from the flat $\S=\mathrm{const}$
dependence, previously suggested\cite{Kormendy:04,Donato:09,Gentile:09a}. The
latter is based, in our view, on a confusion between the properties of
isolated and non-isolated halos. Excellent agreement with pure DM simulations
suggest also that the observed scaling dependence is insensitive to the
presence of baryons, details of local environment, formation history.

The relation~(\ref{eq:2}) can be used to search for deviations from CDM
model~(e.g.  warm DM models\cite{Bode:00}) or modifications of gravity at
large scales\cite{DGP}.  This motivates dedicated astronomical observations
with all the data processed in a uniform way. Studies of the galaxies with the
masses below $10^{10} M_\odot$ and galaxy clusters would be especially
important.

Various scaling relations are known in astrophysics (``fundamental plane
relation'' for elliptical galaxies\cite{Kormendy:89}, ``Tully-Fisher
relation'' for spiral galaxies\cite{Tully:77}, etc.).  The relation discussed
in this Letter differs in one crucial aspect: \emph{it extends uniformly to
  all classes of objects at which DM is observed}. It would be very difficult
to explain such a relation within Modified Newtonian dynamics\cite{Milgrom:83}
theory considered as an alternative to DM.  That is why this relation, further
confirmed, studied and understood analytically, may serve as one more evidence
of the existence of Dark Matter.

If DM particle posses a radiative \emph{decay} channel (as predicted by
several particle physics models), a possible signal in X-ray or $\gamma$-ray
observations would be proportional to the DM column density averaged within
the instrument's field-of-view.  It then follows from our analysis than
decaying DM would produce a unique all-sky signal, with a known slow-varying
angular distribution.  Such a signal can be easily distinguished from any
possible astrophysical background and therefore makes the astrophysical search
for decaying DM \emph{another type of a direct detection
  experiment}\cite{Boyarsky:06c,Bertone:07,Boyarsky:09a}.

\begin{figure} %
  \centering
  \includegraphics[width=\linewidth]{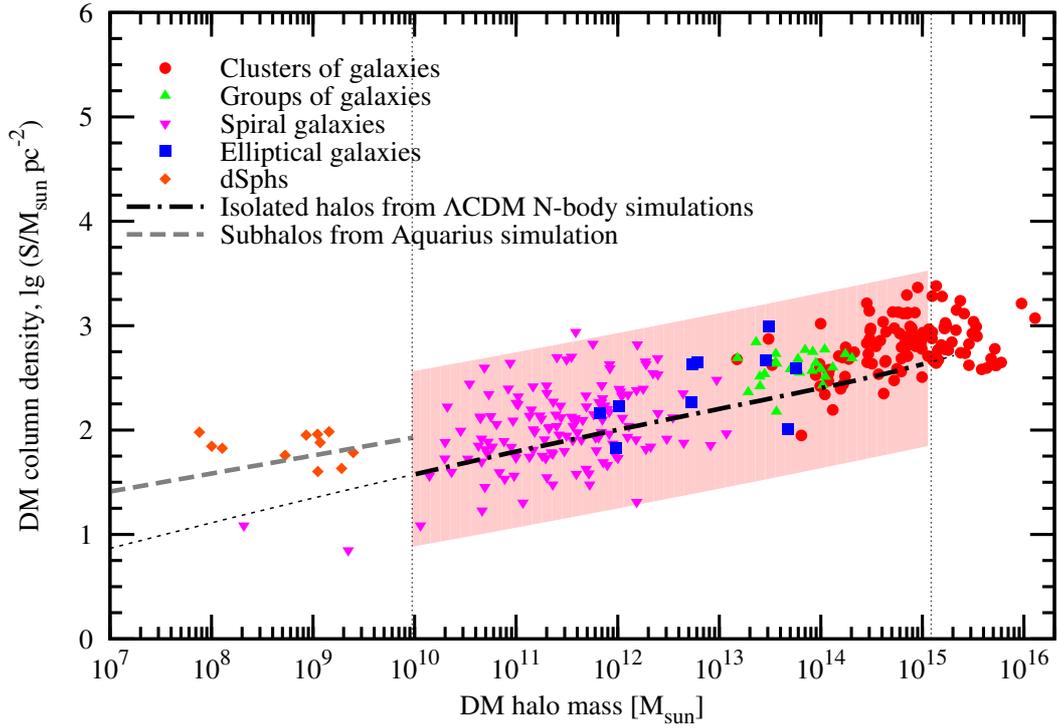} %
  \caption{\textbf{Column density $\S$ as a function of halo mass
      $M_\text{halo}$.}  The black dashed-dotted line is the $S-M_\text{halo}$
    relation obtained from N-body simulations\protect\cite{Maccio:08}, using
    the WMAP fifth year cosmological parameters\cite{WMAP5cosmoParams}. The
    shaded region shows the $3\sigma$ scatter in the simulation data.  The
    vertical lines indicate the mass range probed by simulations.  The dotted
    line is the theoretical prediction from the toy model for isolated
    halos\cite{Bullock:99,Maccio:08}.  The gray dashed line showns the results
    from the Aquarius simulation for satellite halos\cite{Springel:08a}.}
  \label{fig:300M200} 
\end{figure}

\subsubsection*{Method summary}

 We collected from the literature \totalprof DM density profiles for \total
 unique objects ranging from dSphs to galaxy clusters. For each DM profile
 found in our sample we have applied uniform selection criteria:
 \begin{itemize}[--]

\item If for an object several independently determined profiles were
  available and all of them but one agreed in the values of $r_\star$ and $\rho_\star$
  within a factor of 5, we rejected the outlier.

\item For some objects the best-fit value of the characteristic radius
  $r_\star$ was extrapolated well outside the region covered by the
  observational data, $R_\text{data}$. In this case the parameters of the
  density profile had extremely large uncertainties.  We have thus rejected
  objects with $r_\star < 2.75R_\text{data}$.

\item We rejected profiles for which the uncertainty in any quoted parameter
  ($r_\star$ or $\rho_\star$) was higher than a factor of 10.

\item For objects with more than one profile selected, the average value of
  $\S$ and $M_\text{halo}$ was used in the subsequent analysis.

\item 
When processing the data of N-body simulations we used the fit of particle
distribution by the NFW density profile and computed $\S$, using
equation~(\ref{eq:Sbar}).

\item If \emph{the same} observational data is fit by several different DM profiles
(e.g. NFW, ISO, and BURK), one can then find a relation between characteristic
scales $r_\star$ and densities $\rho_\star$ of these profiles. Provided such a
relation holds, the difference between the column densities $\S_\nfw$,
$\S_\burk$ and $\S_\iso$ turns out to be less than $10\%$.  Qualitatively,
this can be understood as follows: to explain the same velocity data, two DM
profiles should have roughly the same mass within some radius $R_0$.  If both
profiles happen to have the same behaviour at large distances, their $\S$
values, averaged over $R_0$ will be essentially equal.  This explains the use
of $\S$ as a characteristic of DM halos.
\end{itemize}

Supplementary information is presented in the Appendix.

\textbf{Acknowledgments.} Numerical simulations were performed on the PIA and
on the PanStarrs2 clusters of the Max-Planck-Institut f\"ur Astronomie at the
Rechenzentrum in Garching.  D.~I. is grateful to to Scientific and Educational
Centre of the Bogolyubov Institute for Theoretical Physics in Kiev, Ukraine,
and especially to V.~Shadura, for creating wonderful atmosphere for young
Ukrainian scientists.  The work of D.~I. is supported from the
``Cosmomicrophysics'' programme, the Program of Fundamental Research of the
Physics and Astronomy Division of the National Academy of Sciences of Ukraine,
and from the grant No.~F/16-417-2007 of the Governmental Fund of Fundamental
Research of the National Academy of Sciences of Ukraine. The work of O.R. was
supported in part by the Swiss National Science Foundation.

\textbf{Author contribution.} A.B. and O.R. suggested the project, contributed
to analysing and interpreting the data, comparing them with numerical
simulations, and writing the paper; D.I. and D.M. collected the data on DM
distributions and contributed to interpreting it; A.M. performed numerical
simulations, contributed to comparison of their results wit the data, and
contributed to writing the paper. Correspondence should be addressed to A.B.
or O.R.  (\textbf{alexey.boyarsky},\textbf{oleg.ruchayskiy@epfl.ch}).

\appendix

\section*{Supplementary Information}

\renewcommand{\figurename}{Supplementary Figure}
\renewcommand{\tablename}{Supplementary Table}

In this Supplementary Information Section we demonstrate that the average DM
column density $\S$ is independent on the choice of the particular DM density
profile; provide more details about the sample that was used and its
comparison with simulations.

\section{Relation between parameters of DM density profiles }
\label{sec:relat-betw-dens}

In this work we concentrated on three popular choices for DM density profile
$\rho(r)$.  Numerical (N-body) simulations of the cold DM model have shown
that the DM distribution in all relaxed halos can be fitted with the universal
Navarro-Frenk-White (NFW) profile\cite{Navarro:96}:
\begin{equation}
  \rho_\nfw(r) = \frac{\rho_s r_s}{r(1+r/r_s)^2}\label{eq:rho_NFW}
\end{equation}
parametrised by $\rho_s$ and $r_s$. A more useful parametrization is in terms
of the halo mass, $M_{200}$, and the concentration parameter, $c\equiv
R_{200}/r_s$.  Namely, $R_{200}$ is the radius at which the average DM density
is 200 times larger than the critical density of the universe $\rho_{\rm
  crit}$. The halo mass $M_{200}$ is the total mass of DM within this radius.
The variables ($\rho_s$,$r_s$) and ($M_{200},c$) are thus connected as
follows:
\begin{eqnarray}
\label{eq:r-rho-c-mvir}
&& \rho_s=f(c)\rho_{\rm
  crit} \nonumber \\ 
&& r_s=\left(\frac{3M_{200}} {800 \pi \rho_{\rm
      crit} c^3}  \right)^{1/3} \\ 
&& f(c)=\frac{200}{3} \frac{c^3}{\ln(c+1)-c/(c+1)}\nonumber
\end{eqnarray} 
The equations (\ref{eq:r-rho-c-mvir}) allow to determine $\S \propto
r_\star\rho_\star$, knowing halo mass and concentration parameter $M_{200},c$.

The Burkert profile\cite{Burkert:95} has been shown to be successful in
explaining the kinematics of disk systems (e.g.\cite{Gentile:04}):
\begin{equation}
\label{eq:1}
\rho_{\textsc{burk}}(r) = \frac{\rho_Br_B^3}{(r_B+r)(r_B^2+r^2)}.
\end{equation}

Another common parametrization of cored profiles is given by the
pseudo-isothermal profile\cite{Kent:86}
\begin{equation}
\rho_\iso(r) = \frac{\rho_c}{1 + r^2/r_c^2}.
  \label{eq:14}
\end{equation}

The quantity $\S (R)$ can be calculated analytically for all these choice of $\rho(r)$.
For example, for the  pseudo-isothermal profile one obtains:
\begin{equation}
  \S_\iso(R) =\frac{2\pi\rho_c r_c^2}{R^2}\left[\sqrt{R^2+r_c^2}-r_c\right].
  \label{eq:9}
\end{equation}
For the NFW density distribution~(\ref{eq:rho_NFW}):
\begin{equation}{\S}_\nfw(R) = \frac{4\rho_s
    r_s^3}{R^2}\left[\frac{\arctan\sqrt{R^2/r_s^2-1}}{\sqrt{R^2/r_s^2-1}} +
    \log\left(\frac{R}{2r_s}\right)\right]\label{eq:S_NFW}\;.
\end{equation}
Notice, that this expression is real for both $R> r_s$ and $R< r_s$.  The
corresponding expression for the BURK is rather lengthy and not very
illuminating.

\subsection{Dependence of $\S$ on the inner slope of density profile.}
\label{sec:depend-inner-slope}

In order to equally well fit the same rotation curve data, two DM
profiles should have roughly the same mass within some radius $R_0$,
determined by the observational data. If both profiles happen to have the same
behaviour at large distances, their $\S$, averaged over $R_0$ will be
essentially equal (as it is determined by the sum of the masses inside the
sphere $R_0$ and in the outside of the cylinders, where the mass is dominated
by the large $r$ asymptotics).\footnote{We will see below (Eq.~(\ref{eq:21})),
  that this is indeed the case for NFW and Burkert profiles.}
In reality the situation is of course more complicated, one has to take into
account the influence of baryons, the span of radii at which the data exists,
etc.

We conservatively estimate the difference of column densities between a cusped
and a cored profile as follows. We take the NFW density
profile~(\ref{eq:rho_NFW}) as a representative of the cusped profile and its
``extreme cored'' counterpart $\rho_{core}(r)$ defined as follows:
\begin{equation}
  \label{eq:10}
  \rho_{core}(r) =\left\{
    \begin{array}{ll}
      \rho_\nfw(r), & r>R_0\\
      \rho_\nfw(R_0), & r\le R_0
    \end{array}\right.\;.
\end{equation}
The column densities of these two profiles, averaged within $R_0$, differ only
because 
\begin{figure}
  \centering
  \includegraphics[width=.75\linewidth]{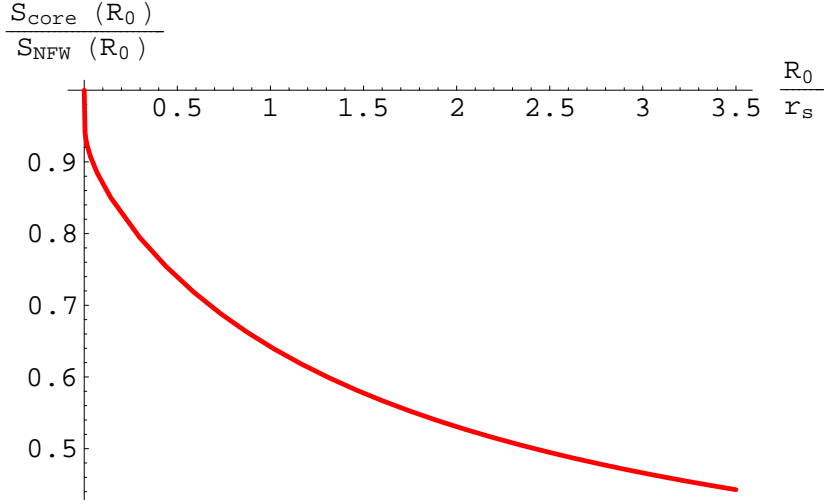}
  \caption{\textbf{The ratio of average column densities of the extreme cored
      and NFW profiles as a function of $R_0$} (c.f. Eq.~(\ref{eq:10})).}
  \label{fig:extreme_core}
\end{figure}
the initial mass inside a \emph{sphere} with radius $R_0$ for the cored
profile~(\ref{eq:10}) diminishes as compared to the NFW case.

The resulting ratio of DM column densities is shown in the Supplementary
Fig.~\ref{fig:extreme_core} as a function of averaging radius $R_0$.  In
particular, for $R_0 = r_s$ this ratio is $64\%$, for $R_0 = 2r_s$ it equals
to $53\%$ and for $R_0 = 3r_s$ it drops to $47\%$.  This implies that the
difference of DM column densities between the cusped (NFW) and the extreme
version of the cored profile~(\ref{eq:10}) is within 50\% for realistic
averaging radii $R_0$ (usually $R_0 \sim 1-3 r_s$).  This difference is small
compared to the intrinsic scatter expected on a object by object basis and
well below the observational uncertainties on the parameters describing the
density profile.  This makes $\S$ a very robust quantity to compare observed
properties of DM halos and results from numerical simulations and,
consequently, test the prediction of the CDM model.

\begin{figure*}
\centering
\begin{tabular}[c]{cc}
\includegraphics[width=.49\linewidth]{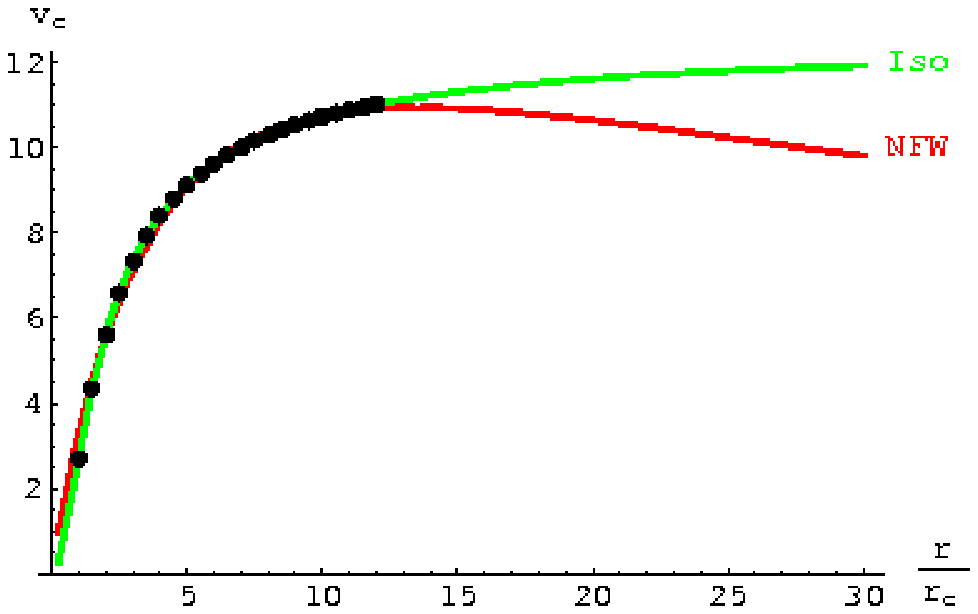} &
\includegraphics[width=.49\linewidth]{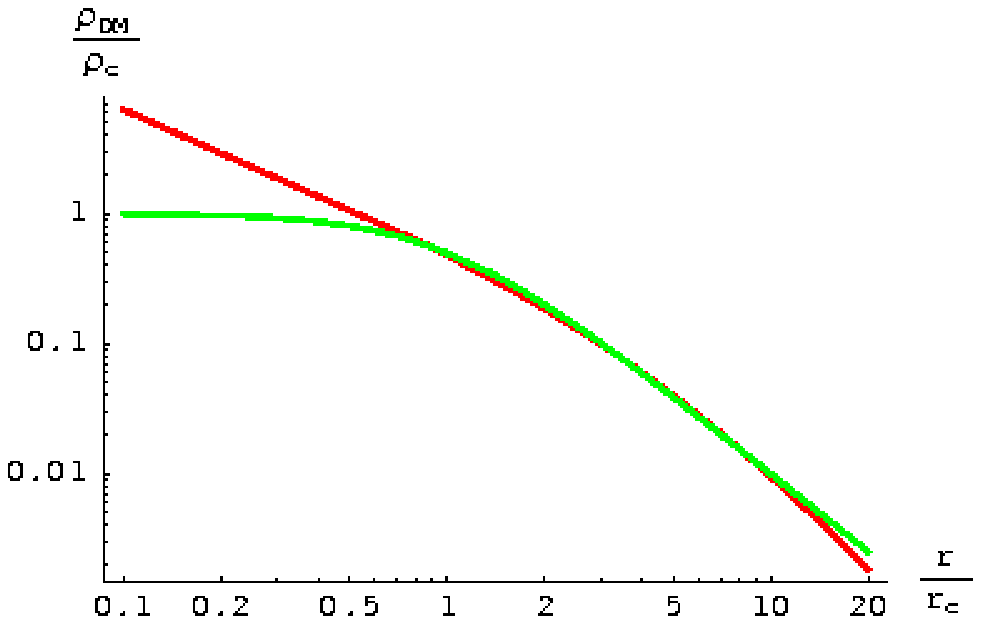}
\end{tabular}
\caption{\textbf{Comparison of NFW (red) and ISO (green) profiles for the
    \emph{simulated} rotation velocity points.} \emph{Left panel:} the
  velocity data (black points, in units of $G_N \rho_c r_c^2$) is generated,
  assuming the ISO profile and fitted with the NFW profile. The parameters of
  the corresponding NFW profile (in the units of $r_c,\rho_c$) are given by
  eq.~(\ref{eq:4}) in the text. \emph{Right panel:} comparison of the density
  profiles with parameters, related by (\ref{eq:4}).  The $x$-axis is in the
  units of isothermal core radius $r_c$.  }
\label{fig:iso_vs_nfw}
\end{figure*}

The rotation curve of a galaxy is often fitted by several DM profiles (e.g.
ISO and NFW).  Let us analytically establish the relation between parameters
of several profiles, \emph{fitting the same rotation curve}.  To this we take
an ISO density profile and generate according to it the circular velocity
profile $v_c^2(r)$, with $r$ in the range  $r_c \lesssim r \lesssim
15r_c$.\footnote{The final result is not sensitive to the exact choice of this
  range.}  Then we fit these data using an NFW profile (see
Fig.~\ref{fig:iso_vs_nfw}, left).
We find the following relations between the parameters of the two profiles:
\begin{equation}
\label{eq:4}
\text{NFW vs. ISO} \quad : \quad r_s \simeq 6.1 \,r_c\quad;\quad \rho_s \simeq
0.11\,\rho_c \;.
\end{equation}
The corresponding rotation curves and density profiles are shown in
Supplementary Fig.~\ref{fig:iso_vs_nfw}.

\begin{figure}
\centering
\includegraphics[width=.75\linewidth]{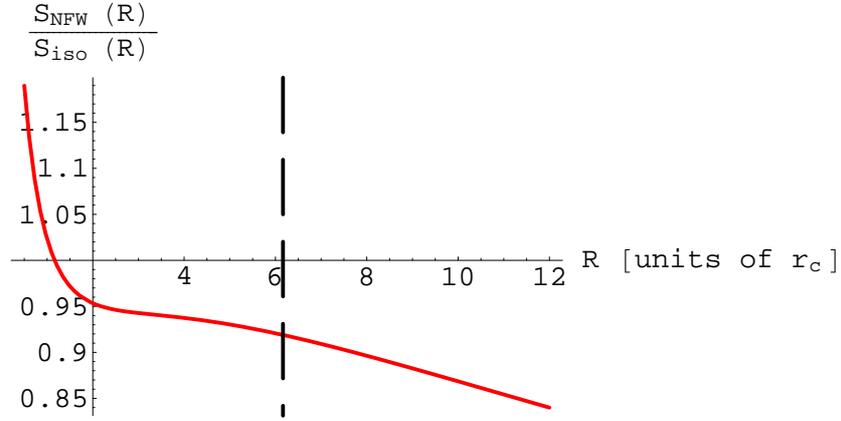} %
\caption{\textbf{Comparison of the column densities of NFW and ISO profiles.}
  Profiles describe the same data and their parameters are related
  via~(\ref{eq:4}). The column density is averaged within various radii $R$.
  Dashed vertical line marks $R=r_s = 6.1r_c$.}
\label{fig:S_iso_nfw}
\end{figure}

Let us now compare the column densities for NFW and ISO profiles, whose
parameters are related via Eq.~(\ref{eq:4}). Results as a function of
radius $R$ are shown on Fig.~\ref{fig:S_iso_nfw}. In particular, one sees that
for $R=r_s$ 
\begin{equation}
\label{eq:15}
\frac{\S_\nfw(r_s)}{\S_\iso(6r_c)} \approx 0.91 \quad .
\end{equation}
One may be surprised that the cusped profile leads to the smaller column
density than the cored one (as Eq.~(\ref{eq:15}) demonstrate).  This result
however, can be simply understood.  We match the velocity profiles for the NFW
and ISO at some off-center distances $R_0 \sim 2 r_s$, by demanding that the
mass inside this sphere is the same for both profiles.  The ISO profile is
shallower in the outer regions than the NFW one.  The ratio between the mass
inside a sphere of the radius $R_0$ and a cylinder with base radius $R_0$ is
equal to $0.58$ at $R_0 = 6r_c$ for ISO profile, while it is $0.63$ at
$R_0=r_s$ for the NFW profile . Thus the mass in the outer part of a cylinder
is larger for the shallower ISO profile than for the cuspy NFW one, which
explains the result~(\ref{eq:15}).

It is clear from previous considerations that $\S_\nfw$ and $\S_{\burk}$
(similarly matched) should be essentially identical, as both profiles have
identical behaviour at $r\rightarrow \infty$. Indeed, in the case of the NFW
and Burkert profiles the relation between their characteristic parameters is
given by
\begin{equation}
\label{eq:5}
\text{NFW vs. BURK} \quad : \quad r_s \simeq 1.6 r_B \quad ; \quad \rho_s \simeq 0.37\rho_B
\end{equation}
which leads to
\begin{equation}
\label{eq:21}
\frac{\S_\nfw(r_s)}{\S_{\burk}(1.6r_s)}\approx 0.98\quad.
\end{equation}
\begin{figure}
\centering
\includegraphics[width=\linewidth]{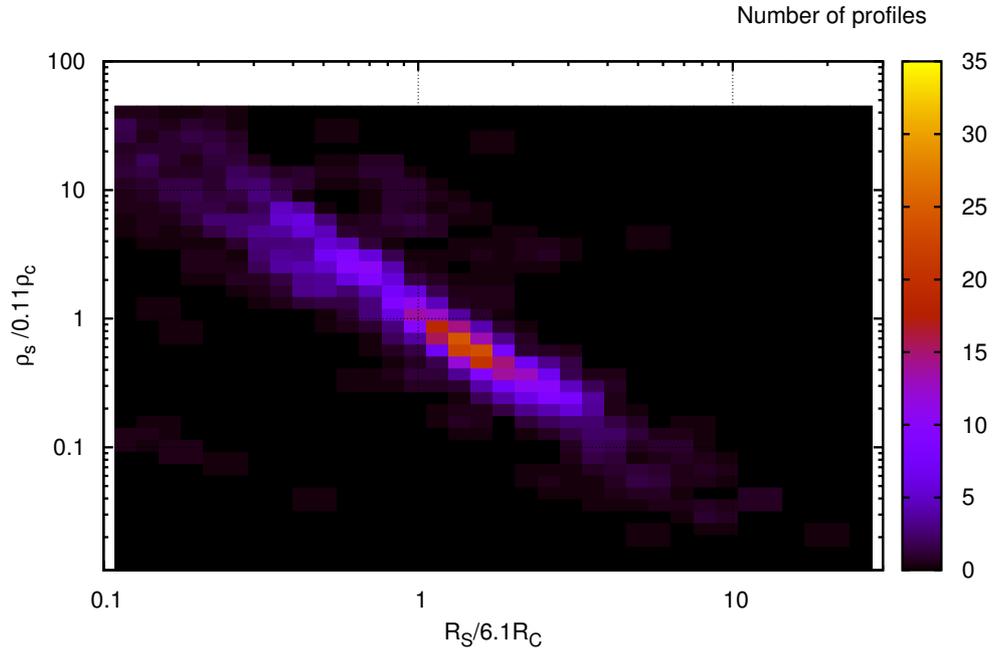}
\caption{\textbf{Relation between parameters of NFW and ISO profiles in
    observed objects.}  For objects for which both NFW and ISO fits of
  velocity rotation curves were available, we plot the ratios $r_s/r_c$ and
  $\rho_s/\rho_c$. The maximum of the histogram lies in a
  region~(\ref{eq:4}).}\label{fig:best_profiles}
\end{figure}

Finally, it should be noticed that we assume an infinite extension for Dark
Matter halos, when computing the column density.  However, the integrals
in~(\ref{eq:1}) are convergent at large off-center distances and therefore
the details of the truncation of the DM distributions for $R>R_{200}$ do not affect the
value of $\S$ by more than 10\%.

\section{Data analysis}
\label{sec:data-selection}

We have collected from the literature \totalprof\ DM profiles for \total\
objects (from dwarf spheroidal galaxies to galaxy clusters, see
Table~\ref{tab:objects} below).  For each DM profile in our sample we have
performed a number of checks.  Those profiles have not passed these checks
were rejected from subsequent analysis. As a result of the selection process
we were left with \totalprofsel\ DM profiles for \totalsel\ objects.  The
results of the selection are shown on the Supplementary
Figure~\ref{fig:only_good} below.

\begin{itemize}[--]
\item When analysing the data, we realized that for some objects the value of
  $r_\star$ lies well outside the region covered by the observational
  data, $R_\text{data}$.  Such objects systematically show extremely high values of
  $r_\star$.  For example, we found 37 galaxy profiles having $r_\star > 100$~kpc,
  while their kinematic data usually extends only up to $R_\text{data} \sim
  10-30$~kpc.  

  Therefore, we select only Dark Matter profiles having $r_\star < 2.75R_\text{data}$.  
  The coefficient $2.75$ is justified by the following argument.  The circular velocity in an NFW
  halo is given by
  \begin{equation}
    v^2_c(r) = \frac{G_N M_{\nfw}(<r)}{r} = 4\pi G_N\rho_s
    r_s^3\frac{\log(1+\frac {r}{r_s})-\frac r{r+r_s}}{r}.\label{eq:3}
  \end{equation}
  For $r \ll r_s$ this function can be approximated as
  \begin{equation}
    \label{eq:7}
    v^2_c(r) \approx  2\pi G_N\rho_s r_s\left( r - \frac{4 r^2}{3 r_s}+\ldots\right)
  \end{equation}
  In the part of the velocity curve where $\frac{4 r^2}{3 r_s}$ is much less
  than the errors on the velocity dispersion one cannot reliably determine
  $r_s$ and $\rho_s$ (since $v_c^2(r)$ is indistinguishable from a straight
  line, proportional to $\rho_s r_s$). It is important to have data points in
  the region where the contribution of the quadratic term becomes noticeable
  to reliably extract both NFW parameters.  We chose to set $2.75
  R_\text{data} \ge r_\star$, which corresponds to a $\sim 50\%$ contribution
  from the second (quadratic) term to $v^2_c(r)$.  Similar criteria are used
  for ISO and BURK profiles. This reduces the number of considered profiles
  from \totalprof to \rmaxsel.

\item For \nfwiso objects both NFW and ISO (or BURK) Dark Matter profiles were
  available. For these objects we checked the relation between the parameters
  of these profiles against the results shown in Eq.~(\ref{eq:4}) (or Eq.
  (\ref{eq:5}) for BURK). Results of this comparison for the NFW and ISO
  profiles are shown in Fig.~\ref{fig:best_profiles}. This figure shows that
  there is indeed a maximum in the region defined by Eq.~(\ref{eq:4}) but also
  that the scatter around this maximum is pretty large and that the difference
  between measured and expected ratios of NFW and ISO parameters can be as
  large as a factor of ten.  Therefore we decided to exclude from our sample
  all objects with a ratio $\rho_s/\rho_c$, $r_s/r_c$, (or $\rho_s/\rho_B$,
  $r_s/r_B$ for BURK profiles) larger than a factor 5 with respect to the
  theoretical prediction shown in Eq.~(\ref{eq:4}) or~(\ref{eq:5}).

\item Finally, in several cases parameters of DM density profiles were quoted
  with very large uncertainties. We decided to select only those profiles for
  which the ratio between the $1\sigma$ upper and lower bounds of quoted
  parameters (radius $r_\star$ or the density $\rho_\star$) {was} 
  smaller than a factor of 10.
\end{itemize}

\begin{figure}
  \centering
  \includegraphics[width=\linewidth]{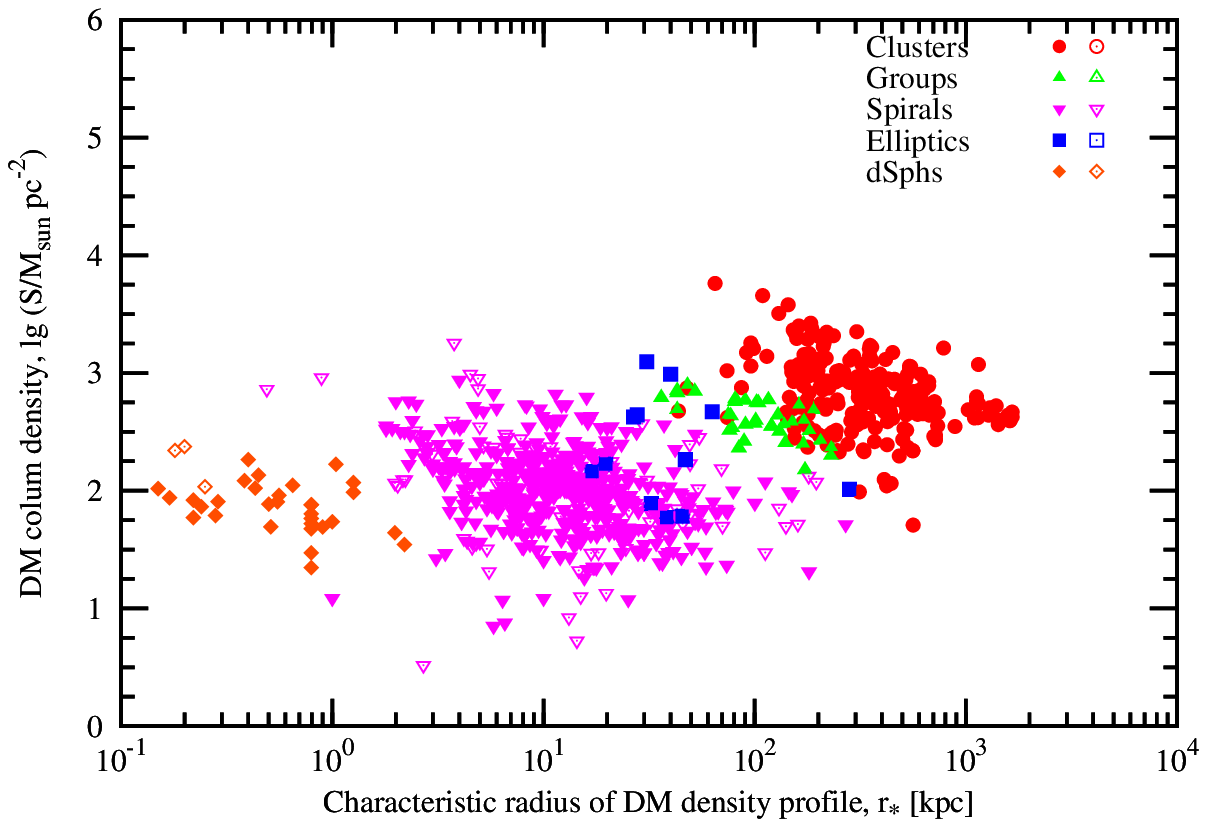}
  \caption{\textbf{Selected DM density profiles.} DM column density $\S$ as a
    function of the characteristic radius $r_\star$ for \totalprof
    \emph{density profiles}. Filled (open) shapes denote accepted (rejected)
    density profiles, according to our selection criteria.  Notice, that
    unlike Figure~1, an object may appear several times in this Figure if more
    than one profile was available. }
\label{fig:only_good}
\end{figure}

To compare the $\S-M_\text{halo}$ relation for selected objects with N-body
simulations, we used the results from\cite{Maccio:08}. This suit of
$\Lambda$CDM numerical simulations probed the halo mass range $10^{10} -
10^{15} M_\odot$.  For each simulated halo of\cite{Maccio:08} we computed
$M_\text{halo}$, fit the particle distribution to the NFW density profile and
calculate $\S$ using Eqs.(\ref{eq:r-rho-c-mvir}) and the
definition~(\ref{eq:Sbar}). The observational data together with results from
simulations is plotted of the Supplementary Figure~\ref{fig:maccio-med}. The
small scatter of the simulation points at $M_\text{halo}\gtrsim
10^{14}M_\odot$ is explained by the finite size of the simulation box. The
simulations with the large box size (e.g. \cite{Neto:07}) verify that the
scatter does not reduce at large masses (c.f. the pink shaded region on the
Figure~1).

\begin{figure}
  \centering
  \includegraphics[width=\linewidth]{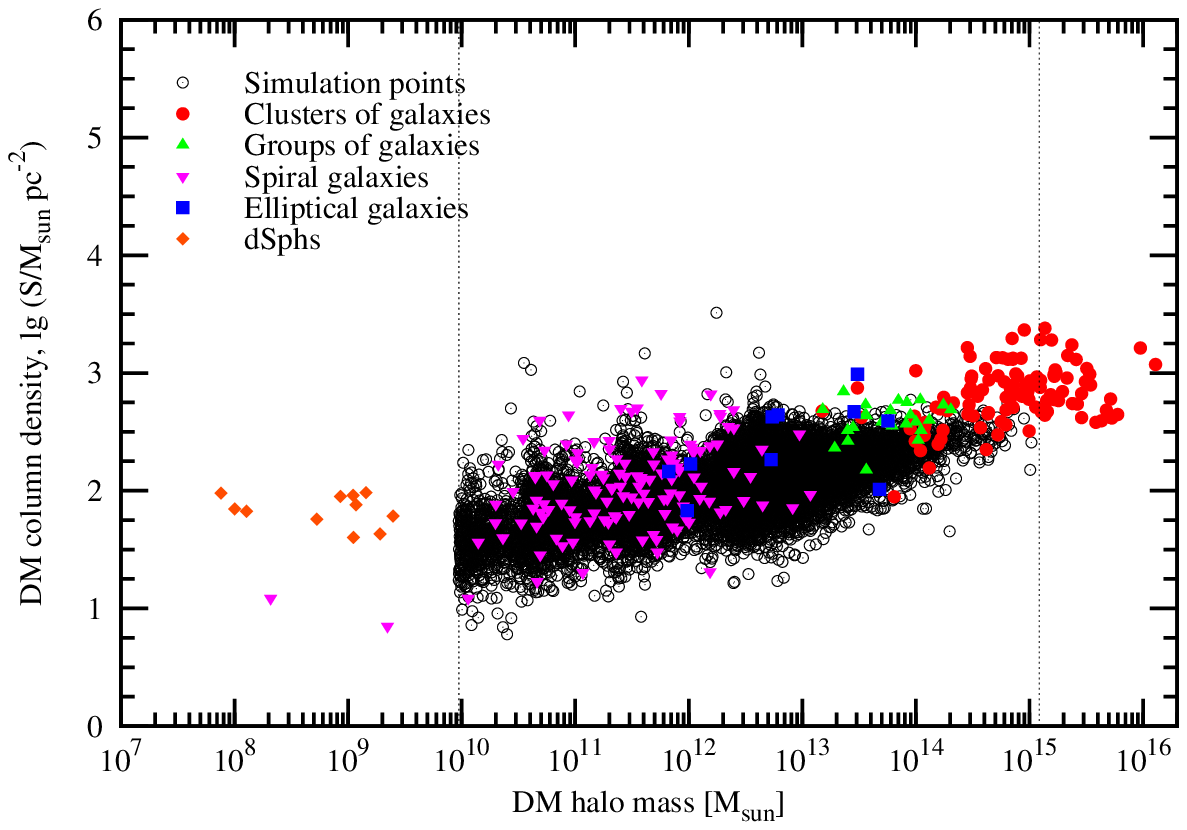} 
  \caption{\textbf{DM Column density as a function of the halo mass.} Similar
    to the Figure~1, we plot \totalsel\ objects, selected in
    Section~\ref{sec:data-selection} above (coloured shapes) superimposed on
    the simulation data for isolated halos\cite{Maccio:08} (open black
    circles).}
  \label{fig:maccio-med}
\end{figure}

\begin{table*}
  \centering %
  \begin{tabular}{lcll}
    Type & Number of objects & References & List of selected objects \\
    \hline
    Galaxy clusters & \clusters & %
    \cite{Bardeau:07,Broadhurst:08,Comerford:07,Corless:08,Donnarumma:09,Ettori:02,%
      Gavazzi:09,Kubo:07,Mclaughlin:98,Rines:03,Schmidt:07,Sereno:09,Umetsu:07} %
    & Table~\ref{tab:clusters}
    \\
    Galaxy groups & \groups & %
    \cite{Gastaldello:06,Romanowsky:08,Sun:08}
    & Table~\ref{tab:groups}
    \\
    Elliptical galaxies & \ellipticals & 
    \cite{Humphrey:06,Weijmans:07,Weijmans:09,Napolitano:08,Zhang:06}
    & Table~\ref{tab:ellipticals} \\
    Spiral galaxies & \spirals & 
    \cite{Athanassoula:87,Battaglia:05,Begeman:91,Begum:04,Blais-Ouellette:99,Blok:02,Blok:08,Blok:97,Bosch:01,Chemin:06,Cote:00,Cote:91,
      Dutton:03,Geehan:05,
      Gentile:04,Kent:86,Kent:87,Lake:89,Lake:90,Marchesini:02,Naray:06,
      Naray:07,Puche:91,Sicotte:97,Spano:07,Swaters:02,Tempel:07,Weldrake:02,Frigerio:09} 
    & Table~\ref{tab:spirals} \\
    Dwarf spheroidal galaxies & \dsphs & 
    \cite{Gilmore:07a,Strigari:07,Wu:07,Walker:09,Frigerio:09,Eymeren:09}%
    & Table~\ref{tab:dwarves}\\
    \hline
    total & \total
  \end{tabular}
  \caption{\textbf{Observational data.} The table lists the types of objects;
    references used to collect the observational data; 
    and the final list of selected objects.}
  \label{tab:objects}
\end{table*}


\footnotesize
\begin{longtable}{|c|c|c|c|c|c|c|c|}
  \caption{Halo mass $M_{200}$ and DM column density $\S$ for galaxy
    clusters.}\label{tab:clusters}\\
  \hline
  Object name & $M_{200} [M_\odot]$ & $\log_{10}\left[\frac{\S}{M_\odot \pc^{-2}}\right]$ &  Object name & $M_{200} [M_\odot]$ & $\log_{10}\left[\frac{\S}{M_\odot \pc^{-2}}\right]$ \\
  \hline
\endfirsthead %
\hline %
Object name & $M_{200} [M_\odot]$ & $\log_{10}\left[\frac{\S}{M_\odot \pc^{-2}}\right]$ & Object name & $M_{200} [M_\odot]$ & $\log_{10}\left[\frac{\S}{M_\odot \pc^{-2}}\right]$ \\
\hline
\endhead
\hline %
\multicolumn{8}{|c|}{{Clusters (\emph{continued on next page})}}\\ %
\hline
\endfoot
\hline %
\multicolumn{8}{|c|}{{Clusters (\emph{end})}} \\ %
\hline
\endlastfoot
\hline %
2A 0335+096 & 1.44e+15 & 2.67 & Abell 1068 & 5.64e+14 & 2.59 \\
Abell 119 & 2.64e+15 & 2.74 & Abell 133 & 4.35e+14 & 2.66 \\
Abell 1367 & 6.89e+14 & 2.72 & Abell 1413 & 9.87e+14 & 2.78 \\
Abell 160 & 9.73e+13 & 2.63 & Abell 1656 (Coma) & 1.72e+15 & 3.01 \\
Abell 168 & 3.19e+14 & 2.64 & Abell 1689 & 1.56e+15 & 3.28 \\
Abell 1703 & 1.25e+15 & 3.28 & Abell 1763 & 9.92e+14 & 2.51 \\
Abell 1795 & 1.76e+15 & 2.79 & Abell 1835 & 1.27e+15 & 2.68 \\
Abell 194 & 8.90e+13 & 2.53 & Abell 1983 & 1.57e+14 & 2.39 \\
Abell 1991 & 1.64e+14 & 2.69 & Abell 2029 & 3.45e+15 & 2.90 \\
Abell 209 & 5.26e+14 & 2.47 & Abell 2142 & 5.98e+15 & 2.65 \\
Abell 2199 & 9.57e+14 & 2.81 & Abell 2204 & 7.24e+14 & 2.90 \\
Abell 2218 & 7.88e+14 & 2.98 & Abell 2219 & 1.65e+15 & 2.77 \\
Abell 2256 & 3.84e+15 & 2.58 & Abell 2319 & 4.75e+15 & 2.68 \\
Abell 2390 & 1.54e+15 & 2.82 & Abell 2537 & 8.18e+14 & 2.82 \\
Abell 2597 & 2.94e+14 & 2.73 & Abell 262 & 1.89e+14 & 2.68 \\
Abell 2667 & 1.37e+15 & 2.64 & Abell 267 & 1.74e+14 & 2.51 \\
Abell 2717 & 1.66e+14 & 2.44 & Abell 3112 & 2.84e+14 & 2.83 \\
Abell 3266 & 5.17e+15 & 2.62 & Abell 3376 & 2.13e+14 & 2.75 \\
Abell 3526 & 9.84e+14 & 2.87 & Abell 3562 & 9.33e+14 & 2.86 \\
Abell 3571 & 5.17e+15 & 2.78 & Abell 3627 & 1.43e+15 & 2.77 \\
Abell 370 & 1.25e+15 & 2.88 & Abell 383 & 6.19e+14 & 2.56 \\
Abell 426 & 1.66e+15 & 2.97 & Abell 478 & 1.42e+15 & 2.76 \\
Abell 496 & 1.64e+15 & 2.99 & Abell 511 & 9.21e+14 & 2.90 \\
Abell 520 & 1.29e+16 & 3.07 & Abell 539 & 2.83e+14 & 3.22 \\
Abell 576 & 6.47e+14 & 3.12 & Abell 611 & 7.43e+14 & 2.82 \\
Abell 644 & 6.96e+14 & 2.72 & Abell 68 & 3.71e+14 & 2.54 \\
Abell 754 & 5.32e+15 & 2.62 & Abell 780 & 9.96e+13 & 3.02 \\
Abell 85 & 2.38e+15 & 2.74 & Abell 851 & 9.85e+13 & 2.42 \\
Abell 907 & 6.24e+14 & 2.79 & Abell 963 & 6.78e+14 & 2.87 \\
Abell S1101 & 1.76e+14 & 2.79 & Abell S400 & 3.20e+15 & 3.04 \\
Abell S540 & 1.69e+14 & 2.71 & AC 114 & 1.27e+15 & 2.67 \\
AWM 4 & 1.50e+14 & 2.71 & CL 0016+1609 & 1.31e+14 & 2.19 \\
CL 2244-0221 & 4.30e+14 & 2.65 & CL J1226.9+3332 & 7.04e+14 & 3.29 \\
B2002a 15 & 2.87e+14 & 2.65 & MACS J0011.7-1523 & 1.14e+15 & 2.71 \\
MACS J0159.8-0849 & 1.25e+15 & 2.94 & MACS J0242.6-2132 & 5.28e+14 & 2.98 \\
MACS J0329.7-0212 & 7.19e+14 & 2.86 & MACS J0429.6-0253 & 4.11e+14 & 3.04 \\
MACS J0744.9+3927 & 1.01e+15 & 2.93 & MACS J0947.2+7623 & 1.16e+15 & 2.97 \\
MACS J1115.8+0129 & 4.32e+15 & 2.59 & MACS J1311.0-0311 & 6.72e+14 & 2.82 \\
MACS J1411.3+5212 & 7.16e+14 & 3.13 & MACS J1423.8+2404 & 5.79e+14 & 3.13 \\
MACS J1427.6-2521 & 3.07e+14 & 2.95 & MACS J1532.9+3021 & 9.40e+14 & 2.86 \\
MACS J1621.6+3810 & 7.58e+14 & 2.99 & MACS J1720.3+3536 & 9.67e+14 & 2.85 \\
MACS J1931.8-2635 & 1.67e+15 & 2.78 & MACS J2229.8-2756 & 3.10e+14 & 2.97 \\
MKW 3S & 1.10e+14 & 2.34 & MKW 9 & 1.19e+14 & 2.54 \\
MS 0015.9+1609 & 9.50e+15 & 3.21 & MS 0116.3-0115 & 1.18e+14 & 2.47 \\
MS 0302.7+1658 & 8.45e+14 & 3.12 & MS 0451.6-0305 & 1.68e+15 & 3.03 \\
MS 0735.6+7421 & 2.17e+15 & 3.15 & MS 0839.9+2938 & 5.99e+14 & 2.93 \\
MS 1006.0+1202 & 3.10e+15 & 2.93 & MS 1008.1-1224 & 3.39e+15 & 2.99 \\
MS 1137.5+6625 & 6.21e+14 & 2.69 & MS 1224.7+2007 & 9.01e+14 & 3.37 \\
MS 1358.4+6245 & 2.58e+15 & 3.12 & MS 1455.0+2232 & 1.37e+15 & 3.38 \\
MS 1512.4+3647 & 7.11e+14 & 3.12 & MS 1910.5+6736 & 8.59e+14 & 2.80 \\
MS 2137.3-2353 & 5.11e+14 & 3.13 & PKS 0745-19 & 2.14e+15 & 2.96 \\
PKS 0745-191 & 2.87e+15 & 2.82 & RXC J0137.2-0911 & 9.83e+13 & 2.61 \\
RXC J0454.8-1806 & 1.21e+14 & 2.58 & RXC J1504.1-0248 & 1.95e+15 & 2.84 \\
RX J0056.9-2740 & 6.45e+13 & 1.95 & RX J0419.6+0225 & 3.06e+13 & 2.87 \\
RX J0439.0+0520 & 4.39e+14 & 2.94 & RX J0439.0+0521 & 3.85e+14 & 2.86 \\
RX J1023.6+0411 & 2.88e+15 & 2.62 & RX J1223.0+1037 & 3.31e+13 & 2.62 \\
RX J1347.5-1145 & 2.36e+15 & 3.24 & RX J1416.4+2315 & 3.00e+14 & 3.14 \\
RX J1504.1-0248 & 1.80e+15 & 2.77 & RX J2129.6+0005 & 7.15e+14 & 2.71 \\
SDSS J1004+4112 & 3.02e+14 & 2.87 & Triangulum & 3.32e+15 & 2.91 \\
Virgo & 4.16e+14 & 2.35 & ZWCL 0024.0+1652 & 3.49e+14 & 2.80 \\
ZWCL 1321.4+1358 & 1.50e+13 & 2.68 & & & \\
\end{longtable}

\begin{longtable}{|c|c|c|c|c|c|c|c|}
  \caption{Halo mass $M_{200}$ and DM column density $\S$  for galaxy
    groups.\label{tab:groups}}\\ %

  \hline
  Object name & $M_{200} [M_\odot]$ & $\log_{10}\left[\frac{\S}{M_\odot \pc^{-2}}\right]$ &  Object name & $M_{200} [M_\odot]$ & $\log_{10}\left[\frac{\S}{M_\odot \pc^{-2}}\right]$ \\
  \hline
\endfirsthead %
\hline %
Object name & $M_{200} [M_\odot]$ & $\log_{10}\left[\frac{\S}{M_\odot \pc^{-2}}\right]$ &  Object name & $M_{200} [M_\odot]$ & $\log_{10}\left[\frac{\S}{M_\odot \pc^{-2}}\right]$ \\
\hline
\endhead
\hline %
\multicolumn{8}{|c|}{{Groups (\emph{continued on next page})}}\\ %
\hline
\endfoot
\hline %
\multicolumn{8}{|c|}{{Galaxy Groups }} \\ %
\hline
\endlastfoot
\hline %
Abell 1275 & 8.20e+13 & 2.57 & Abell 1692 & 1.09e+14 & 2.77 \\
Abell 2462 & 1.11e+14 & 2.51 & Abell 2550 & 8.93e+13 & 2.64 \\
Abell 3880 & 2.03e+14 & 2.69 & ESO 306-G017 (group) & 1.74e+14 & 2.73 \\
ESO 351-G021 (group) & 3.66e+13 & 2.18 & ESO 552-G020 (group) & 1.31e+14 & 2.60 \\
IC 1860 (group) & 5.02e+13 & 2.58 & MKW 4 & 5.97e+13 & 2.68 \\
NGC 1407 (group) & 6.13e+13 & 2.52 & NGC 1550 (group) & 3.61e+13 & 2.73 \\
NGC 2563 (group) & 2.51e+13 & 2.51 & NGC 4104 (group) & 6.18e+13 & 2.55 \\
NGC 4325 (group) & 3.57e+13 & 2.64 & NGC 5044 (group) & 3.67e+13 & 2.64 \\
NGC 5098 (group) & 2.52e+13 & 2.42 & NGC 5129 (group) & 1.51e+13 & 2.69 \\
NGC 533 (group) & 2.31e+13 & 2.84 & RGH 80 & 2.80e+13 & 2.53 \\
RX J1022.1+3830 & 1.00e+14 & 2.58 & RX J1158.1+5521 & 1.05e+14 & 2.43 \\
RX J1159.8+5531 & 8.33e+13 & 2.75 & UGC 5088 (group) & 1.93e+13 & 2.36 \\
UGC 842 (group) & 7.00e+13 & 2.77 & & & \\
\hline
\end{longtable}

\begin{longtable}{|c|c|c|c|c|c|c|c|}

  \caption{Halo mass $M_{200}$ and DM column density $\S$ for elliptical
    galaxies.\label{tab:ellipticals}}\\%
  \hline
  Object name & $M_{200} [M_\odot]$ & $\log_{10}\left[\frac{\S}{M_\odot \pc^{-2}}\right]$ &  Object name & $M_{200} [M_\odot]$ & $\log_{10}\left[\frac{\S}{M_\odot \pc^{-2}}\right]$ \\
  \hline
\endfirsthead %
\hline %
Object name & $M_{200} [M_\odot]$ & $\log_{10}\left[\frac{\S}{M_\odot \pc^{-2}}\right]$ &  Object name & $M_{200} [M_\odot]$ & $\log_{10}\left[\frac{\S}{M_\odot \pc^{-2}}\right]$ \\
\hline
\endhead

\hline %
\multicolumn{8}{|c|}{{Elliptical galaxies (\emph{continued on next page})}}\\ %
\hline
\endfoot

\hline %
\multicolumn{8}{|c|}{{Elliptical galaxies }} \\ %
\hline
\endlastfoot

\hline %
M49 (NGC 4472) & 2.85e+13 & 2.67 & M60 (NGC 4649) & 3.05e+13 & 2.99 \\
NGC 4125 & 5.32e+12 & 2.26 & NGC 4261 & 4.77e+13 & 2.01 \\
NGC 6482 & 6.12e+12 & 2.64 & NGC 720 & 5.43e+12 & 2.63 \\
 \hline
\end{longtable}

\begin{longtable}{|c|c|c|c|c|c|c|c|}
  \caption{Halo mass $M_{200}$ and DM column density $\S$ for dwarf spheroidal
    galaxies (dSphs).\label{tab:dwarves}}\\ 
  \hline
  Object name & $M_{200} [M_\odot]$ & $\log_{10}\left[\frac{\S}{M_\odot \pc^{-2}}\right]$ &  Object name & $M_{200} [M_\odot]$ & $\log_{10}\left[\frac{\S}{M_\odot \pc^{-2}}\right]$ \\
  \hline
\endfirsthead %
\hline %
Object name & $M_{200} [M_\odot]$ & $\log_{10}\left[\frac{\S}{M_\odot \pc^{-2}}\right]$ &  Object name & $M_{200} [M_\odot]$ & $\log_{10}\left[\frac{\S}{M_\odot \pc^{-2}}\right]$ \\
\hline
\endhead

\hline %
\multicolumn{8}{|c|}{{Dwarf spheriodal galaxies (\emph{continued on next page})}}\\ %
\hline
\endfoot

\hline %
\multicolumn{8}{|c|}{{Dwarf spheriodal galaxies }} \\ %
\hline
\endlastfoot

\hline %
Carina & 2.63e+09 & 1.64 & Coma Berenices & 1.00e+08 & 1.85 \\
Draco & 1.25e+09 & 1.97 & Fornax & 3.04e+09 & 1.54 \\
Leo II & 7.17e+08 & 1.69 & Ursa Major II & 1.27e+08 & 1.83 \\
Ursa Minor & 1.85e+09 & 2.01 & Willman I & 7.61e+07 & 1.98 \\

\end{longtable}

\newpage

\begin{longtable}{|c|c|c|c|c|c|c|c|}
  \caption{Halo mass $M_{200}$ and DM column density $\S$ for spiral,
    low-surface brightness
    (LSB) and dwarf galaxies (without dSphs).}\label{tab:spirals}\\

  \hline
  Object name & $M_{200} [M_\odot]$ & $\log_{10}\left[\frac{\S}{M_\odot \pc^{-2}}\right]$ &  Object name & $M_{200} [M_\odot]$ & $\log_{10}\left[\frac{\S}{M_\odot \pc^{-2}}\right]$ \\
  \hline
\endfirsthead %
\hline %
Object name & $M_{200} [M_\odot]$ & $\log_{10}\left[\frac{\S}{M_\odot \pc^{-2}}\right]$ & Object name & $M_{200} [M_\odot]$ & $\log_{10}\left[\frac{\S}{M_\odot \pc^{-2}}\right]$ \\
\hline
\endhead
\hline %
\multicolumn{8}{|c|}{{Spiral, LSB and dwarf galaxies (without dSphs) (\emph{continued on next page})}}\\ %
\hline
\endfoot
\hline %
\multicolumn{8}{|c|}{{Spiral, LSB and dwarf galaxies (without dSphs)
    (\emph{end})}} \\ %
\hline
\endlastfoot
\hline %
  DDO 154 & 5.52e+10 & 1.44 & DDO 170 & 6.82e+10 & 1.60 \\
DDO 189 & 4.26e+10 & 1.85 & DDO 47 & 2.13e+11 & 2.03 \\
DDO 52 & 2.00e+10 & 1.89 & DDO 64 & 5.28e+10 & 1.89 \\
ESO 116-G12 & 5.14e+10 & 1.65 & ESO 287-G13 & 1.36e+09 & 0.95 \\
ESO 79-G14 & 4.41e+08 & 1.08 & IC 1401 & 1.54e+12 & 1.31 \\
IC 2574 & 2.28e+11 & 1.59 & KK 98250 & 4.57e+10 & 1.70 \\
LSBC F563-1 & 2.78e+11 & 1.95 & LSBC F563-v2 & 2.41e+11 & 2.10 \\
LSBC F568-1 & 4.35e+11 & 2.28 & LSBC F568-3 & 4.66e+11 & 1.89 \\
LSBC F568-v1 & 2.37e+11 & 2.25 & LSBC F571-8 & 5.74e+11 & 2.25 \\
LSBC F574-1 & 2.15e+11 & 2.05 & LSBC F583-1 & 2.56e+11 & 1.75 \\
LSBC F583-4 & 9.37e+10 & 1.82 & M31 & 8.30e+11 & 2.63 \\
M33 (NGC 598) & 6.48e+11 & 1.93 & Milky Way & 1.03e+12 & 2.40 \\
NGC 100 & 2.50e+11 & 2.04 & NGC 1003 & 5.30e+11 & 1.48 \\
NGC 1085 & 4.37e+12 & 2.35 & NGC 1087 & 7.06e+11 & 2.10 \\
NGC 1090 & 3.01e+08 & 1.06 & NGC 1353 & 1.27e+12 & 2.34 \\
NGC 1421 & 1.60e+12 & 1.93 & NGC 1560 & 7.91e+11 & 1.67 \\
NGC 224 & 1.18e+13 & 1.97 & NGC 2366 & 2.51e+10 & 1.79 \\
NGC 2403 & 3.80e+11 & 2.11 & NGC 247 & 4.47e+12 & 1.88 \\
NGC 2532 & 4.39e+11 & 2.18 & NGC 2537 & 7.85e+10 & 2.29 \\
NGC 2552 & 1.44e+12 & 2.38 & NGC 2608 & 3.10e+11 & 1.94 \\
NGC 2708 & 1.86e+12 & 2.63 & NGC 2841 & 1.87e+12 & 2.66 \\
NGC 2903 & 6.96e+11 & 2.43 & NGC 2977 & 4.54e+11 & 2.47 \\
NGC 300 & 8.95e+11 & 1.87 & NGC 3026 & 1.39e+12 & 2.43 \\
NGC 3031 & 2.52e+11 & 2.70 & NGC 3109 & 1.16e+11 & 1.30 \\
NGC 3198 & 3.64e+11 & 2.04 & NGC 3200 & 2.16e+12 & 2.53 \\
NGC 3274 & 5.50e+10 & 2.46 & NGC 3319 & 8.11e+11 & 1.74 \\
NGC 3346 & 4.04e+11 & 2.50 & NGC 3495 & 1.68e+13 & 3.33 \\
NGC 3521 & 5.80e+11 & 2.15 & NGC 3621 & 6.05e+11 & 1.93 \\
NGC 3741 & 1.99e+10 & 1.73 & NGC 3893 & 1.98e+12 & 2.84 \\
NGC 3898 & 9.42e+12 & 2.48 & NGC 4190 & 5.20e+11 & 1.59 \\
NGC 4236 & 1.95e+11 & 1.73 & NGC 4242 & 1.08e+12 & 1.80 \\
NGC 4258 & 3.04e+12 & 1.92 & NGC 4288 & 1.00e+11 & 2.26 \\
NGC 4395 & 6.40e+10 & 2.06 & NGC 4455 & 1.00e+11 & 1.78 \\
NGC 4494 & 8.65e+11 & 1.87 & NGC 45 & 8.25e+11 & 1.97 \\
NGC 4605 & 1.06e+11 & 2.33 & NGC 4635 & 5.21e+11 & 2.53 \\
NGC 4736 & 2.30e+11 & 2.48 & NGC 4800 & 3.51e+11 & 2.70 \\
NGC 5023 & 1.35e+11 & 2.14 & NGC 5033 & 1.25e+12 & 1.93 \\
NGC 5055 & 2.17e+12 & 1.86 & NGC 5194 & 1.56e+12 & 2.82 \\
NGC 5204 & 9.53e+10 & 2.15 & NGC 55 & 1.00e+12 & 1.73 \\
NGC 5585 & 3.40e+11 & 1.85 & NGC 5608 & 4.97e+10 & 1.68 \\
NGC 5727 & 5.41e+12 & 2.06 & NGC 5949 & 1.72e+11 & 2.33 \\
NGC 5963 & 6.08e+11 & 1.76 & NGC 6015 & 1.25e+12 & 2.49 \\
NGC 6503 & 6.90e+11 & 1.98 & NGC 6682 & 5.22e+10 & 1.78 \\
NGC 6689 & 2.76e+11 & 2.12 & NGC 6946 & 6.19e+12 & 2.06 \\
NGC 7137 & 4.45e+10 & 2.12 & NGC 7217 & 2.47e+12 & 2.69 \\
NGC 7339 & 3.54e+11 & 1.89 & NGC 753 & 1.92e+12 & 1.82 \\
NGC 7541 & 2.52e+12 & 2.17 & NGC 7664 & 1.26e+12 & 2.03 \\
NGC 7793 & 3.68e+11 & 2.16 & NGC 925 & 6.16e+11 & 1.81 \\
NGC 959 & 4.87e+10 & 2.60 & UGC 02259 & 9.49e+11 & 1.76 \\
UGC 10310 & 1.81e+11 & 1.71 & UGC 11557 & 4.47e+11 & 1.84 \\
UGC 11707 & 2.11e+11 & 2.03 & UGC 11820 & 1.99e+11 & 2.43 \\
UGC 11861 & 8.14e+11 & 2.21 & UGC 12060 & 5.37e+11 & 1.98 \\
UGC 1230 & 1.47e+11 & 2.01 & UGC 12632 & 4.55e+10 & 1.92 \\
UGC 12732 & 1.47e+11 & 1.81 & UGC 128 & 3.84e+11 & 2.08 \\
UGC 1281 & 9.96e+10 & 1.74 & UGC 12810 & 1.78e+12 & 2.40 \\
UGC 1551 & 4.98e+10 & 1.82 & UGC 2259 & 5.37e+10 & 2.35 \\
UGC 2885 & 8.22e+12 & 1.86 & UGC 3137 & 2.10e+11 & 2.02 \\
UGC 3371 & 1.81e+11 & 1.75 & UGC 4173 & 4.62e+10 & 1.23 \\
UGC 4499 & 1.03e+11 & 1.85 & UGC 477 & 4.00e+11 & 1.98 \\
UGC 5005 & 4.52e+11 & 1.54 & UGC 5750 & 2.50e+11 & 1.46 \\
UGC 628 & 2.73e+11 & 2.23 & UGC 6446 & 5.83e+10 & 1.84 \\
UGC 731 & 4.42e+10 & 2.13 & UGC 7699 & 1.44e+11 & 1.70 \\
UGC 9211 & 2.84e+10 & 1.99 & & & \\

\end{longtable}

\normalsize

\let\jnlstyle=\rm\def\jref#1{{\jnlstyle#1}}\def\aj{\jref{AJ}}
\def\araa{\jref{ARA\&A}} \def\apj{\jref{ApJ}\ } \def\apjl{\jref{ApJ}\ }
\def\apjs{\jref{ApJS}} \def\ao{\jref{Appl.~Opt.}} \def\apss{\jref{Ap\&SS}}
\def\aap{\jref{A\&A}} \def\aapr{\jref{A\&A~Rev.}} \def\aaps{\jref{A\&AS}}
\def\azh{\jref{AZh}} \def\baas{\jref{BAAS}} \def\jrasc{\jref{JRASC}}
\def\memras{\jref{MmRAS}} \def\mnras{\jref{MNRAS}\ }
\def\pra{\jref{Phys.~Rev.~A}\ } \def\prb{\jref{Phys.~Rev.~B}\ }
\def\prc{\jref{Phys.~Rev.~C}\ } \def\prd{\jref{Phys.~Rev.~D}\ }
\def\pre{\jref{Phys.~Rev.~E}} \def\prl{\jref{Phys.~Rev.~Lett.}}
\def\pasp{\jref{PASP}} \def\pasj{\jref{PASJ}} \def\qjras{\jref{QJRAS}}
\def\skytel{\jref{S\&T}} \def\solphys{\jref{Sol.~Phys.}}
\def\sovast{\jref{Soviet~Ast.}} \def\ssr{\jref{Space~Sci.~Rev.}}
\def\zap{\jref{ZAp}} \def\nat{\jref{Nature}\ } \def\iaucirc{\jref{IAU~Circ.}}
\def\aplett{\jref{Astrophys.~Lett.}}
\def\apspr{\jref{Astrophys.~Space~Phys.~Res.}}
\def\bain{\jref{Bull.~Astron.~Inst.~Netherlands}}
\def\fcp{\jref{Fund.~Cosmic~Phys.}} \def\gca{\jref{Geochim.~Cosmochim.~Acta}}
\def\grl{\jref{Geophys.~Res.~Lett.}} \def\jcp{\jref{J.~Chem.~Phys.}}
\def\jgr{\jref{J.~Geophys.~Res.}}
\def\jqsrt{\jref{J.~Quant.~Spec.~Radiat.~Transf.}}
\def\memsai{\jref{Mem.~Soc.~Astron.~Italiana}}
\def\nphysa{\jref{Nucl.~Phys.~A}} \def\physrep{\jref{Phys.~Rep.}}
\def\physscr{\jref{Phys.~Scr}} \def\planss{\jref{Planet.~Space~Sci.}}
\def\procspie{\jref{Proc.~SPIE}} \let\astap=\aap \let\apjlett=\apjl
\let\apjsupp=\apjs \let\applopt=\ao

\end{document}